\documentclass[aps,prb,twocolumn,showpacs,superscriptaddress,groupedaddress]{revtex4-1}
\usepackage{graphicx}
\usepackage{amsmath}
\usepackage{amssymb,color}
\usepackage[utf8]{inputenc}
\usepackage{braket}
\input{epsf}

\newcommand{\nn}{\nonumber}
\newcommand{\kk}{\textbf{k}}
\newcommand{\qv}{\textbf{q}}
\newcommand{\rv}{\textbf{r}}
\newcommand{\Rv}{\textbf{R}}

\newcommand{\so}{\text{sgn}(\omega)}

\begin{document}

	\title{Phase-sensitive determination of nodal $\mathbf{d}$-wave order parameter in single-band and multiband superconductors}
	
	\author{Jakob~B\"oker$^1$}
	\author{Miguel Antonio Sulangi$^2$}
	\author{Alireza Akbari$^{3,4}$}
	\author{J.C. Séamus Davis$^{5,6,7}$}	
	\author{P. J. Hirschfeld$^2$}
	\author{Ilya M. Eremin$^1$}

	\affiliation{1-Institut f\"ur Theoretische Physik III, Ruhr-Universit\"at Bochum, D-44801 Bochum, Germany}
	\affiliation{2-Department of Physics, University of Florida, Gainesville, Florida 32611, USA}
	\affiliation{3-Max Planck Institute for the Chemical Physics of Solids, D-01187 Dresden, Germany}
	\affiliation{4-Max Planck POSTECH Center for Complex Phase Materials, POSTECH, Pohang 790-784, Korea}
	\affiliation{5-LASSP, Department of Physics, Cornell University, Ithaca, New York 14850, USA }
	\affiliation{6-Clarendon Laboratory, University of Oxford, Oxford, OX1 3PU, United Kingdom}
	\affiliation{7-Department of Physics, University College Cork, Cork T12 R5C, Ireland}
	
	\begin{abstract}
		Determining the exact pairing symmetry of the superconducting order parameter in candidate unconventional superconductors remains an important challenge. Recently, a new method, based on phase sensitive quasiparticle interference measurements, was developed to identify gap sign changes in isotropic multiband systems. Here we extend this approach to the single-band and multiband nodal $d$-wave superconducting cases relevant,  respectively, for the cuprates and likely for the infinite-layer nickelate superconductors. Combining analytical and numerical calculations, we show that the antisymmetrized correction to the tunneling density of states due to nonmagnetic impurities in the Born limit and at intermediate-scattering strength shows characteristic features for sign-changing and sign-preserving scattering wave vectors, as well as for the  momentum-integrated quantity. Furthermore, using a realistic approach accounting for the Wannier orbitals, we model scanning tunneling microscopy data of $\text{Bi}_2\text{Sr}_2\text{CaCu}_2\text{O}_{8+\delta}$, which should allow the comparison of our theory with experimental data.
	\end{abstract}
	\date{\today}
	
	\maketitle
	
	\section{Introduction}
	
	The determination of the symmetry properties of the superconducting order parameter in unconventional superconductors is an important step towards an understanding of the underlying Cooper-pairing mechanism in these systems. Therefore phase-sensitive experimental techniques based on observables that can be routinely measured are very much in demand. One rapidly developing technique makes use of quasiparticle interference (QPI) as measured by Fourier transformed scanning tunneling microscopy (STM) [\onlinecite{hoffman2002imaging,mcelroy2003relating,kohsaka2008cooper,lee2009spectroscopic,hanaguri2009coherence,hoffman2011spectroscopic,fujita2014simultaneous}]. This probe measures the wavelength of Friedel oscillations caused by impurities in a metallic or superconducting system, which in turn contain information on the electronic structure of the pure system [\onlinecite{wang2003quasiparticle, capriotti2003wave, zhu2004power, nunner2006fourier, pereg2008magnetic, maltseva2009model, vishik2009momentum, kreisel2015interpretation, dalla2016holographic, sulangi2017revisiting}]. Recently, this technique became particularly powerful in the iron-based superconductors where conventional phase-sensitive methods could not be easily applied [\onlinecite{Mazin2009, allan2012anisotropic, allan2013anisotropic, allan2015identifying, kostin2018imaging}]. It was proposed by Hirschfeld, Altenfeld, Eremin and Mazin (HAEM) [\onlinecite{Hirschfeld2015}] that the sign structure of the superconducting order parameter in a multiband system can be extracted from the Fourier-transformed local density of states (LDOS) QPI pattern near a weak nonmagnetic impurity. This so-called HAEM method states that the antisymmetrized LDOS $\rho^-(\omega)$, when integrated over wave vectors corresponding to scattering between two bands with order parameters $\Delta_1\neq\Delta_2$, has a dependence on frequency very different for gap sign-changing and sign-preserving  processes $\text{sgn}(\Delta_1)=\mp\text{sgn}(\Delta_2)$. While in the former case $\rho^-(\omega)$ exhibits a strong single-sign enhancement without a sign change of the signal between the two gap scales, the intensity in the latter is strongly suppressed and does change sign between $\Delta_1$ and $\Delta_2$. We refer to this as \emph{even} and \emph{odd} behavior in $\rho^-(\omega)$, respectively.
	
	The HAEM method has mostly been tested on putative isotropic gap states in iron-based superconductors [\onlinecite{sprau2017discovery,Du2018,Cheung2020}]; the limitations of the method have been checked for multiple-impurity scattering [\onlinecite{Martiny2017}] and more complex band structures [\onlinecite{AltenfeldLiFeAs2018}]. Furthermore, inspired by the HAEM method its extension with respect to the impurity bound state has been recently proposed [\onlinecite{Chi2017a}] and verified independently by experiment [\onlinecite{Chen2019}, \onlinecite{Gu2019}].
	
	In principle, the HAEM method should work for anisotropic nodal sign-changing gaps in single-band and multiband systems as well. This includes the canonical $d$-wave state in cuprates and potentially the newly discovered  infinite-layer nickelate superconductors [\onlinecite{Li2019_Ni_Discovery}]. Nevertheless, it is not {\it a priori} clear how these distinctions, developed for the case of two mostly isotropic gaps on different bands, apply to the case of a sign-changing gap on a single band. In the original approach [\onlinecite{Hirschfeld2015},\onlinecite{AltenfeldLiFeAs2018}], the \emph{odd} or \emph{even} behavior of $\rho^-(\omega)$ occurs between the two gap scales, whereas in the $d$-wave case there is only one gap scale present. However, one can think of the $d$-wave system as containing a distribution of gap edges as one goes around the Fermi surface. Consequently, each scattering process which connects $\kk$ and $\kk+\qv$ will reproduce the HAEM phenomenology in its contribution to $\rho^-(\omega)$, according to whether there is a sign change between $\Delta_\kk$ and $\Delta_{\kk+\qv}$.
	
	In this manuscript, we show that the general predictions of HAEM continue to hold and allow an unambiguous determination of a sign change in the gap not merely in a fully momentum-integrated fashion, but also in a {\it momentum-resolved way}. In particular, we show that the conclusions hold individually for a set of octet vectors $\qv_{i}$ connecting the tips of contours of constant quasiparticle energy. To do so, we calculate  $\rho^-_d(\qv,\omega)$, assuming the usual $d$-wave gap $\Delta^d_\kk=\frac{\Delta_0}{2}(\cos k_x-\cos k_y)$, and compare with $\rho^-_s(\qv,\omega)$ resulting from a hypothetical anisotropic $s$-wave state $\Delta^s_\kk=\frac{\Delta_0}{2}|\cos k_x-\cos k_y |$. Both gaps clearly produce the same quasiparticle density of states, but are well distinguished by their phase structure. We discuss our results in the context of a single-band model for the cuprate superconductor Bi$_2$Sr$_2$CaCu$_2$O$_{8+x}$ (BSCCO) and a multiband model for the infinite-layer nickelates.
	
	\section{HAEM for a single-band $\mathbf{d}$-wave superconductor} 
	
	Our starting point is a one-band BCS model Hamiltonian in the Nambu space $H_\kk=\epsilon_{\kk}\tau_3+\Delta_\kk\tau_1$. The corresponding Green's function in the superconducting state and the LDOS are then given by $\hat{G}_\kk(\omega)=[(\omega+i\delta)\tau_0-H_{\kk}]^{-1}$ and
	\begin{align}
	\rho(\omega)=-\frac{1}{\pi}\sum_{\kk} \text{Im}\text{Tr}\frac{(\tau_0+\tau_3)}{2}\hat{G}_\kk(\omega)\label{Eq:LDOS}.
	\end{align}
	Following the HAEM prescription [\onlinecite{Hirschfeld2015},\onlinecite{AltenfeldLiFeAs2018}] the antisymmetrized correction to the LDOS due to impurity scattering in the Born limit is given by the convolution of the bare Green's functions dressed by scattering matrix from a nonmagnetic impurity, $\hat{V}=V_0\tau_3$.  
	\begin{align}
	\delta\rho^-(\qv,\omega)&=\delta\rho(\qv,\omega)-\delta\rho(\qv,-\omega)\label{Eq:Rho Minus}\\
	\delta\rho(\qv,\omega)&=-\frac{1}{\pi}\sum_{\kk}\text{Im}\text{Tr}\frac{(\tau_0+\tau_3)}{2}\hat{G}_\kk(\omega)\hat{V}\hat{G}_{\kk+\qv}(\omega) \label{Eq:CorrextionLDOS}.
	\end{align}
	Before presenting the results for the lattice model, we investigate Eq.~(\ref{Eq:Rho Minus}) analytically using the parabolic band dispersion $\epsilon_\kk=\frac{\kk^2}{2m}-\mu$ and the superconducting order parameters, defined in terms of the spherical harmonics on the Fermi surface, {\it i.e.},  $\Delta^d_\kk=\Delta_0\cos(2\varphi)$ and $\Delta^s_\kk=\Delta_0|\cos(2\varphi)|$, corresponding to the $d$-wave and anisotropic sign-preserving $s$-wave case, respectively. This allows us to approximate $\sum_{\kk}\approx N_0\int_{0}^{2\pi} \frac{d\varphi}{2\pi}\int_{-\infty}^{\infty}d\epsilon_\kk$, which yields the momentum integrated Green's functions as
	\begin{align}
	&\hat{G}^d(\omega)=-2iN_0K\left[\frac{\Delta_0}{\omega}\right]\label{Eq:SimpleGwd}\tau_0,\\
	&\hat{G}^s(\omega)=-2iN_0\Bigg[ K\left[\frac{\Delta_0}{\omega}\right]\tau_0-\nn\\
	&\Bigg(\ln\left(\frac{\sqrt{|\omega^2-\Delta_0^2|}}{\Delta_0+|\omega|}\right)\text{sgn}(\omega)+\frac{i\pi}{2}\theta\left[\Delta_0^2-\omega^2\right]\Bigg)\tau_1\Bigg],\label{Eq_SimpleGws}
	\end{align}
	where $K\left[\frac{\Delta_0}{\omega}\right]$ is the elliptic function of the first kind, see also Appendix~\ref{Sec:MomentumIntegratedGreensfunction} for further details, $N_0=\frac{m}{2\pi}$, and $\theta$ is the Heaviside step function.
	Note that Eq.~($\ref{Eq:SimpleGwd}$) has no $\tau_1$ component since the $d$-wave gap averages to zero when integrating over the polar angle.  
	
\begin{figure}[t]
		\centering
		\includegraphics[width=1\linewidth]{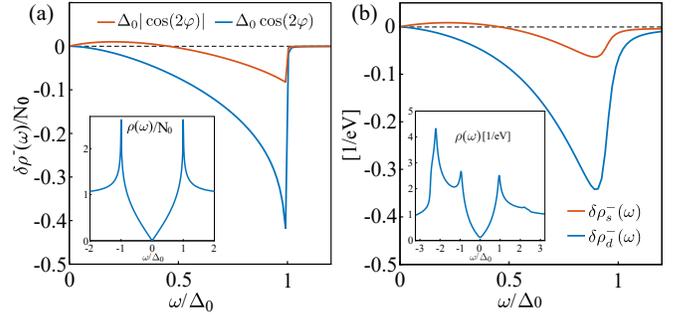}
		\caption{(a) $\delta\rho^-_{d(s)}(\omega)$  calculated using a parabolic band dispersion as a function of $\omega/\Delta_0$ for the sign-changing (blue) and sign-preserving (orange) order parameter. (b) Same quantity as in (a) in units of [states/eV/spin] calculated using the BSCCO lattice model [Eq.~(\ref{Eq:BandDispersion})]. Inset: LDOS $\rho(\omega)$ obtained from the lattice model [Eq.~(\ref{Eq:BandDispersion})] in the superconducting state for either $d$- or $s$-wave state considered here.}
		\label{Fig:DOSSingle$d$-wave}
	\end{figure}

	Both pairing symmetries $\Delta^d_\kk$ and $\Delta^s_\kk$ lead to identical LDOS patterns $\rho(\omega)=\frac{2N_0}{\pi}\text{Im}\enspace iK\left[\frac{\Delta_0}{\omega}\right]$ with the typical V-shape of the nodal superconductor, shown in the inset of Fig.~\ref{Fig:DOSSingle$d$-wave}(a). Using Eqs.~(\ref{Eq:SimpleGwd}) and (\ref{Eq_SimpleGws}), we calculate $\delta\rho^-_{d(s)}(\omega)$ for the $d$- and $s$-wave cases, respectively as shown in  Fig.~\ref{Fig:DOSSingle$d$-wave}(a), where we assume $V_0>0$ without loss of generality.
	In the $d$-wave case, $\delta\rho^-_{d}(\omega)$ is negative for energies between zero and the coherence peak energy where its intensity diverges. In sharp contrast to this, $\delta\rho^-_{s}(\omega)$ is positive at low energies, changes slope and becomes negative upon approaching coherence peak energy where it reaches a much weaker intensity maximum than $\delta\rho^-_{d}(\omega)$. Analytically, for $\omega/\Delta_0\ll1$, we find
	\begin{align}
	\delta\rho^-_d(\omega)&\approx-\text{sgn}(\omega)\frac{8V_0N_0^2}{\pi}\frac{\omega^2}{\Delta_0^2}\ln\left(\frac{4\Delta_0}{|\omega|}\right),\label{drho(i)Approx}\\
	\delta\rho^-_s(\omega)&\approx\frac{8V_0N_0^2}{\pi}\left[\frac{\omega}{\Delta_0}-\text{sgn}(\omega)\frac{\omega^2}{\Delta_0^2}\ln\left(\frac{4\Delta_0}{|\omega|}\right)\right]\label{drho(ii)Approx},
	\end{align}
	where, due to the term linear in $\omega$, $\delta\rho^-_s(\omega)$ is positive at low energies. At $\omega=\Delta_0$,
	\begin{align}
	\delta\rho^-_d(\omega)&=-\so4V_0N_0^2\ln\left(\frac{\pi}{\sqrt{\delta}}\right),\\
	\delta\rho^-_s(\omega)&=-\so4V_0N_0^2\ln\left(\frac{\pi}{2}\right),
	\end{align}
	where $\delta$ is a positive infinitesimal.
	Note that  $\delta\rho^-_s(\omega)$ is now negative but finite, while $\delta\rho^-_d(\omega)$ diverges logarithmically for $\delta\rightarrow0$. This observation is fully consistent with the HAEM prediction for the sign-changing and sign-preserving gaps [\onlinecite{Hirschfeld2015}]. In particular, $\delta\rho^-_d(\omega)$ obeys \emph{even} behavior as a function of frequency between $\omega=0$ and the gap scale $\omega=\Delta_0$, showing a high intensity at the latter. On contrary, $\delta\rho^{-}_s(\omega)$ is \emph{odd} and converges towards a lower intensity.
	
	In the following, we show that our findings for the analytical model are qualitative in nature and hold as well for the lattice model relevant for the cuprates.
	%Furtheremore, the \emph{even} behavior accompanied by high intensity can be clearly attributed to scattering events connecting parts of the Brillouin zone between which the ordering parameter changes sign, while scattering events connecting sign-preserving areas display \emph{odd} behavior and suppressed intensity.
	% 	possibly have an \emph{odd} pattern and low intensity cause.
	To describe the energy dispersion of optimally doped Bi$_2$Sr$_2$CaCu$_2$O$_{8+\delta}$, we employ the following tight-binding parametrization:
	\begin{align}
	\epsilon_\kk=\mu+&\frac{t_1}{2}(\cos{k_x}+\cos{k_y})+t_2\cos{k_x}\cos{k_y}\nn\\
	&+\frac{t_3}{2}(\cos{2k_x}+\cos{2k_y}).\label{Eq:BandDispersion}
	\end{align}
	Here, following Ref.~[\onlinecite{Eschrig2003}] we choose the hopping integrals and the chemical potential as $(t_1,t_2,t_3, \mu)$=(-590.8, 96.2, -130.6, 156.6) meV and 
	$\Delta^{d}_\kk=\frac{\Delta_0}{2}(\cos{k_x}-\cos{k_y})$,  $\Delta^{s}_\kk=\frac{\Delta_0}{2}|\cos{k_x}-\cos{k_y}|$,  with $\Delta_0=31$ meV. The resulting LDOS and   $\delta\rho^-_{d(s)}(\omega)$ are shown in  Fig.~\ref{Fig:DOSSingle$d$-wave}(b). Most importantly, we observe that the qualitative behaviors for the sign-changing and the sign-preserving gaps are quite characteristic and agree well with the results of the analytical calculations shown in Fig.~\ref{Fig:DOSSingle$d$-wave}(a). It reinforces  the fact that the qualitative result of HAEM method also holds in the lattice model despite the presence of sizable particle-hole asymmetry in the normal state.
	
	Note that the results shown in Fig.~\ref{Fig:DOSSingle$d$-wave} for the momentum integrated $\delta\rho^-_{d(s)}(\omega)$ can only be used so far to indicate the overall sign-changing gap structure in the entire Brillouin zone, just as in the case of iron-based superconductors [\onlinecite{AltenfeldLiFeAs2018}]. To make more specific statements about the momentum-dependence of the gap, we need now to make a momentum-dependent analysis, which is done below.
	\begin{figure}[t]
		\centering
		\includegraphics[width=1\linewidth]{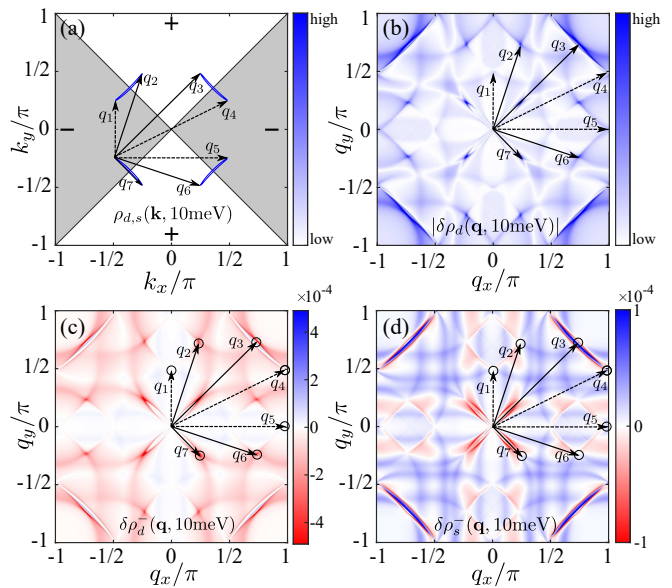}
		\caption{(a) Contours of constant quasiparticle energy  for $\omega=0.3\Delta_0=10$ meV. The octet scattering vectors connecting the tips of the banana-shaped pockets in the superconducting state are shown as $q_1-q_7$. The sign of the $d$-wave gap is indicated by the gray (negative) and white (positive)  regions. (b), (c) and (d) show QPI patterns  $|\delta\rho_{d}(\qv,10\text{ meV})|$, $\delta\rho^-_{d}(\qv,10\text{ meV})$, and $\delta\rho^-_{s}(\qv,10\text{ meV})$, respectively, with the nonmagnetic impurity placed in the center of the image. The circles in (c) and (d) indicate areas of integration around different scattering events. }
		\label{Fig:QPI_Banana_Maps}
	\end{figure}
	It is well-known [\onlinecite{hoffman2002imaging, mcelroy2003relating,wang2003quasiparticle,capriotti2003wave}] that specific scattering wave vectors dominate the QPI spectrum of the $d$-wave superconductor due to the so-called banana-shape of the quasiparticle energy contours below the superconducting gap energy, $\Delta_0$.  In Fig.~\ref{Fig:QPI_Banana_Maps}(a), we show these banana-like contours of constant quasiparticle energy at $10$meV ($\approx 0.3 \Delta_0$) and indicate the characteristic octet vectors $q_1$ to $q_7$ that connect the banana tips. Here, the white and gray colors cover regions where the $d$-wave order parameter is positive and negative, respectively. $q_1$, $q_4$, and $q_5$ connect regions in which the gap has the same sign and we display them as dashed arrows, while $q_2$, $q_3$, $q_6$, and $q_7$, displayed as solid arrows, connect regions involving a sign change of the gap. QPI patterns in the cuprates consist of relatively well-defined spots associated with each of these octet vectors. 
	
	In Fig.~\ref{Fig:QPI_Banana_Maps}(b), we show the $|\delta\rho_d(\qv,\omega)|$ QPI map with the same scattering vectors as in Fig.~\ref{Fig:QPI_Banana_Maps}(a) indicating the positions of the spots from the octet model. Some are oriented exactly along high symmetry lines: the sign-preserving wave vectors $q_1$ and $q_5$ are parallel to $(\pi,0)$, while the sign-changing wave vectors  $q_3$ and $q_7$ are parallel to $(\pi,\pi)$ direction.  In  Figs.~\ref{Fig:QPI_Banana_Maps}(c) and \ref{Fig:QPI_Banana_Maps}(d), we present the antisymmetrized momentum dependent conductance maps $\delta\rho^-_{d/s}(\qv,\omega)$ and the corresponding octet vectors. Observe that the pattern along $(\pi,\pi)$ in  Fig.~\ref{Fig:QPI_Banana_Maps}(c) is more intense than in $(\pi,0)$ direction. Moreover, especially in the vicinity of the octet spots the signal is dominated by strong negative intensities along $(\pi,\pi)$ wave vector, and is weakly positive along $(\pi,0)$ [and $(0,\pi)$] direction. A different situation occurs for the sign-preserving anisotropic $s$-wave gap as can be seen in Fig.~\ref{Fig:QPI_Banana_Maps}(d), where all vectors are sign-preserving. In sharp contrast to its $d$-wave counterpart,  here the signals in $(\pi,\pi)$ and $(\pi,0)$ direction are similarly intense and the octet spots are dominated by positive intensity. Importantly, the intensities at $\mathbf{q}$-vector locations are much smaller in the $s$-wave case than in the $d$-wave case. 
	
\begin{figure}[t]
		\centering
		\includegraphics[width=1\linewidth]{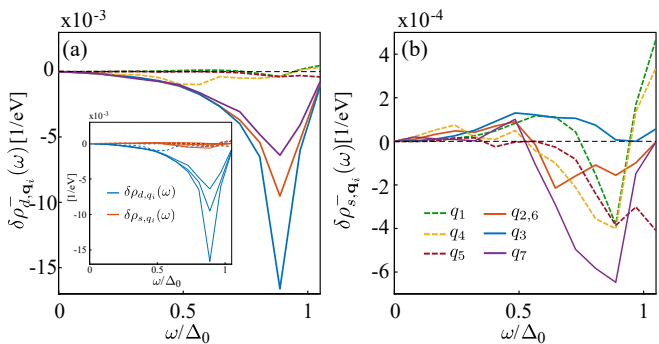}
		\caption{Antisymmetrized correction to LDOS integrated in the vicinity of each octet vector in units of [states/eV/spin] for (a) $d$-wave symmetry and (b) its sign-preserving anisotropic $s$-wave counterpart. Solid (dashed) curves correspond to signals that stem from sign-changing (sign-preserving) octet scattering vectors. (Inset) Results from (a) and (b) plotted in solid/dashed-blue and solid/dashed-orange on the same energy scale. Observe also the larger magnitude of the sign-changing scattering contributions as compared to the sign-preserving ones. }
		\label{Fig:q_integratedII}
	\end{figure}

	In particular, one can analytically calculate $\delta\rho^-(\kk,\qv,\omega)$ ``pointwise'' (\emph{i.e.}, at specific values of $\mathbf{q}$) and evaluate it at the banana tips where $\kk=\kk_F$ and $\qv=\qv_{1-7}$, with $\kk_F$ the Fermi wave vector.  Exactly at these points we have $\epsilon_{\kk_F}=\epsilon_{\kk_F+\qv_i}=0$ and $\Delta_{\kk_F+\qv_i}=\pm\Delta_{\kk_F}$ with ``$+$'' solution for sign-preserving $(\qv_1,\qv_4,\qv_5)$ and ``$-$'' solution for sign-changing $(\qv_{2/6},\qv_3,\qv_7)$. After a straightforward calculation (see appendix~\ref{Suppl:AnalyticalPoints}), we find $\delta\rho^-_{+-}(\omega)/\delta\rho^-_{++}(\omega)\approx -2\Delta_{\kk_F}/\epsilon$, where $\epsilon=\Delta_{\kk_F}-\omega\ll\Delta_{\kk_F}$ and $++/+-$ indices label sign-preserving and sign-changing contributions, respectively.
	We conclude that the HAEM technique can be used to determine which octet vectors correspond to sign-changing scattering processes by integrating over a small range of $\qv$ around each octet spot individually. To do so, we numerically calculate the full momentum-space QPI maps and identify octet vectors for each bias voltage from $|\delta\rho(\qv,\omega)|$ [see Fig.~\ref{Fig:QPI_Banana_Maps}(b)]. Then we compute the antisymmetrized correction to the LDOS, $\delta\rho^-(\qv,\omega)$, and integrate it in the vicinity of each octet vector $\qv_i$ as depicted by black circles in Figs.~\ref{Fig:QPI_Banana_Maps}(c) and \ref{Fig:QPI_Banana_Maps}(d).  Note that there are two reasons why it is useful to integrate the signal within a region around each {\bf q}-spot. First, in FT-STM experiments, {\bf q}-spots have a finite size and generally do not occupy a single pixel. Thus, in analyzing the experiment one typically integrates the experimental data in its vicinity  in order to capture all features from the spot, see, e.g., Refs.[\onlinecite{sprau2017discovery},\onlinecite{Gu2019}]. In addition, observe that  the octet vectors $\textbf{q}_i$ can be measured only up to some accuracy in experiment, thus it is natural to average over some $\delta q$ interval in its vicinity also in theoretical calculations. We do this for both pairing symmetries $\Delta^d_\kk$ and $\Delta^s_\kk$ and present our $\qv_i$-integrated results $\delta\rho^-_{d(s),\qv_i}(\omega)$ in Figs.~\ref{Fig:q_integratedII}(a) and \ref{Fig:q_integratedII}(b).
	The solid curves in Fig.~\ref{Fig:q_integratedII}(a) refer to the sign-changing scattering vectors $q_{2,6}$, $q_{3}$ and $q_7$ and display \emph{even} behavior over the entire energy range $0<\omega\leq \Delta_0$. (Note that vectors $q_2$ and $q_6$ are equivalent via $\pi/2$ rotation.) The dashed curves refer to the sign-preserving scattering vectors $q_1$, $q_{4}$ and $q_5$. Their overall intensity, especially in the vicinity of $\omega=\Delta_0$, is an order of magnitude lower than that for the solid curves and their behavior is \emph{odd} for the frequency interval $0<\omega\leq \Delta_0$. Furthermore, the $q_1$ and $q_5$ contributions, integrated along the $(\pi,0)$ direction, have the lowest intensity.
	
	In Fig.~\ref{Fig:q_integratedII}(b), the calculations are repeated for the sign-preserving $s$-wave case. All curves, with the exception of $q_1$, show \emph{odd} frequency behavior for $0<\omega\leq \Delta_0$ and vary within the same intensity scale. In the inset of Fig.~\ref{Fig:q_integratedII}(a), the results of (a) and (b) are plotted as solid/dashed-blue and solid/dashed-orange curves, respectively, in one plot.  Near the coherence peak at $\omega=\Delta_0$ two characteristic  behaviors can be identified: an intense signal resulting from sign-changing scattering, and a weak signal caused by sign-preserving scattering. We conclude that within the HAEM theory for a weak impurity, sign-changing scattering processes can be  distinguished by the \emph{even} behavior of $\delta\rho^-_{\qv_i}(\omega)$ accompanied by an intense increase of intensity near the coherence peak energy. Scattering events that preserve the sign, however, show a weak intensity for each frequencies and an overall odd frequency dependence in the frequency interval $0 < \omega \leq \Delta_0$.
	
	The results presented so far in the weak-scatterer Born limit for the pointlike impurity are promising, yet their validity has to be tested beyond the Born limit. Furthermore, we need to account for surface wave functions that contribute to the tunneling conductance.  We discuss these extensions in the next two sections.
	
	\section{QPI using continuum Green's function representation}
	
	To compare theoretically calculated QPI images with actual STM experiments, it is important to know the exact location of the STM tip with respect to the Cu atoms. Following Refs. [\onlinecite{Choubey2014},\onlinecite{kreisel2015interpretation}], we do this by calculating the \emph{continuum} Green's function $\hat{G}_{\rv \rv^\prime}(\omega)=\sum_{\Rv \Rv^\prime}\omega_{\Rv}(\rv)\omega^*_{\Rv^\prime}(\rv^\prime)\hat{G}_{\Rv \Rv^\prime}(\omega) $ within a Wannier basis $\omega_{\Rv}(\rv)$ describing the continuum position at $\rv$. Here the lattice Green's function $\hat{G}_{\Rv,\Rv^\prime}(\omega)=\sum_{\kk,\kk^\prime}G_{\kk,\kk^\prime}(\omega)\text{e}^{i\kk\Rv-i\kk^\prime\Rv^\prime}$ is the equivalent of the momentum-space Green's function $\hat{G}_\kk(\omega)$, investigated in the previous sections, but now defined on a real-space lattice. The continuum Green's function method allows us to capture the tunneling contributions from the vacuum layer above the surface where the tip is located, and includes contributions from neighboring $3d_{x^2-y^2}$ orbitals; it thus naturally accounts for $d$-wave-like filter effects [\onlinecite{MartinBalatskyZaanen2002}]. Furthermore, it yields a higher resolution in momentum space due to  its inclusion of intra-unit cell contributions, in contrast to discrete lattice models where the wave-vector size is restricted by the lattice spacing $a$ to lie only within the first Brillouin zone $(\kk,\qv)=(-\frac{\pi}{a},\frac{\pi}{a})$. This lack of resolution in discrete lattice models especially becomes a problem if the length of a scattering vector between two banana tips exceeds $\frac{2\pi}{a}$ (\emph{i.e.}, the full extent of the first Brillouin zone). 
	
	The Wannier function can be obtained numerically by downfolding first-principles calculations [\onlinecite{KuRosnerPicketScalettar2002}], and is primarily of Cu $d_{x^2-y^2}$ character [\onlinecite{kreisel2015interpretation},\onlinecite{dalla2016holographic}]. For the sake of simplicity, we approximate the Wannier function by 
		\begin{align}
		\omega_\textbf{R}(\textbf{r})=\alpha_\omega\left[\left(\frac{x-R_x}{a}\right)^2-\left(\frac{y-R_y}{a}\right)^2\right]\text{e}^{\left(\frac{|\textbf{r}-\textbf{R}|}{r_a}\right)^\gamma}. \label{Eq:WannierFunction}
		\end{align}
	
\begin{figure}[t]
		\centering
		\includegraphics[width=1\linewidth]{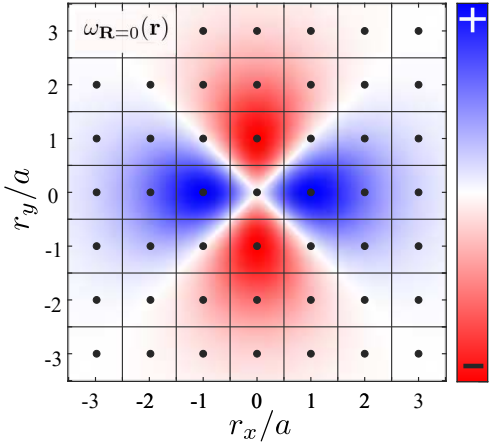}		
		\caption{Spatial dependence of the Wannier function,  Eq.~(\ref{Eq:WannierFunction}), using $\gamma=1.5$ on the CuO$_2$ lattice.}	
		\label{Fig:d_wave_Wannier}
	\end{figure}

\begin{figure}[t]
		\centering
		\includegraphics[width=1\linewidth]{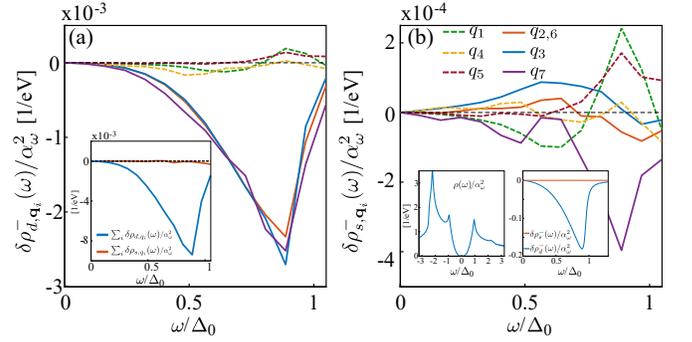}
		\caption{Calculations of the LDOS and $\delta \rho^-({\bf q}_i\omega)/\alpha^2_\omega$  in units of [states/eV/spin] for the $d$-wave (a) and sign-preserving $s$-wave (b) superconducting gaps within a continuum Green's function approach, based on Eqs.~(\ref{Eq:BandDispersion})-(\ref{Eq:ContinuumLDOS}), including a Wannier function approximation. The inset in (a) shows the summed results of (a) and (b) within the same energy scale.  The left and right insets in (b) show the momentum-integrated LDOS $\rho(\omega)$ and the antisymmetrized correction $\delta\rho^-_{d(s)}(\omega)$ obtained from Eqs.~(\ref{Eq:BandDispersion})-(\ref{Eq:ContinuumLDOS}). }
		\label{Fig:q_integratedWannier}
	\end{figure}

	Here, the vector $\Rv=(R_x,R_y)=a(n,m)$ connects the Cu sites, $n,m\in\mathbb{N}_0$, $a$ is the lattice constant and $\rv=(x,y)$.  The parameters $r_a=a(\gamma/2)^{(1/\gamma)}$ and $\gamma$ are chosen such that $\omega_\textbf{R}(\textbf{r})$ decays within a circle of radius $3a$ and has its maximum (minimum) at $x=(\pm a,0)$ ($y=(0, \pm a)$) in agreement with Ref.~[\onlinecite{kreisel2015interpretation}]. The constant normalization factor $\alpha_\omega$ is given in dimensions of $L^{-2/3}$ and its value can, in principle, be estimated by comparison to 
		exact numerical methods mentioned above. Note that Eq.~(\ref{Eq:WannierFunction}) does not strictly obey orthogonality in the sense that $\int d\rv\omega_{\Rv_i}(\rv) \omega_{\Rv_j}(\rv)=\delta_{ij}$. However, it correctly reproduces the spatial dependence of the $d_{x^2-y^2}$ orbital Wannier function and the resulting QPI maps (see Fig.~\ref{Fig:d_wave_Wannier}).
	Using Eq.~(\ref{Eq:WannierFunction}) and $\omega_\textbf{k}(\rv)=\sum_\Rv \omega_\textbf{R}(\rv)\text{e}^{i\kk\Rv}$, the LDOS is given by
	\begin{align}
	\rho(\textbf{r},\omega)=-\frac{1}{\pi}&\text{Im}\text{Tr}\frac{(\tau_0+\tau_3)}{2}\Bigg[\sum_{\textbf{\kk}}|\omega_\kk(\textbf{r})|^2G_{\kk}(\omega)\nn\\
	+&\sum_{\kk,\qv}\omega_\kk(\textbf{r})G_{\kk}(\omega)\hat{V}\omega^*_{\kk+\qv}(\textbf{r})G_{\kk+\qv}(\omega)\Bigg],\label{Eq:ContinuumLDOS}
	\end{align}
	from which $\delta\rho(\qv,\omega)$ in units of states/energy/spin/volume is obtained by Fourier transforming the LDOS map. The momentum-integrated LDOS $\rho(\omega)$ in the Born limit using Eqs.~(\ref{Eq:BandDispersion})-(\ref{Eq:ContinuumLDOS}) is plotted in the left inset of Fig.~\ref{Fig:q_integratedWannier}(b), showing the expected $U$ shape [\onlinecite{kreisel2015interpretation}]. The corresponding $\delta \rho^-(\omega)$ is shown in the right inset Fig.~\ref{Fig:q_integratedWannier}(b) and displays again \textit{even} frequency dependence for the $d$-wave symmetry and \textit{odd} frequency dependence of the sign-preserving $s$-wave, respectively. Analytically, by assuming a parabolic dispersion and using $\omega_\kk(\rv=0)\sim\cos(k_x)-\cos(k_y)$ (which is equivalent to the $d$-wave filter, but one which involves only nearest-neighbor lattice sites), we find $\rho_{d(s)}(\omega)\sim\frac{|\omega|^3}{\Delta_0^3}$, $\delta\rho^-_s(\omega)\approx0$ for all frequencies in the range $0<\omega \leq \Delta_0$, and  $\delta\rho^-_d(\omega)\sim-\so\frac{\omega^2}{\Delta_0^2}$ for $\omega/\Delta_0<<1$, $<0$ and  $\delta\rho^-_d(\omega)\sim-\so\ln\left(\frac{\pi}{\sqrt{\delta}}\right)$ for $\omega=\Delta_0$.
	In Figs.~\ref{Fig:q_integratedWannier}(a) and \ref{Fig:q_integratedWannier}(b), we also present $\delta\rho^-_{d(s),\qv_i}(\omega)$ for $d$- and $s$-wave gap symmetry, respectively.  These results are consistent with those in the previous section.\newline\newline
	We remark once again that the above results were all obtained under the assumption of  weak scatterers. Regarding cuprates, there is a significant amount of data, collected over the last decade from STM showing that various types of dopants behave as weak to intermediate scatterers.   For example, in Refs.~[\onlinecite{kreisel2015interpretation},\onlinecite{sulangi2017revisiting}] it was argued that several QPI features in BSCCO are best explained by weak point-like scattering which is an argument for the presence of weak impurities in BSCCO. In addition, analysis of transport data at optimal and overdoped cuprates, which includes an 
	{\it ab-initio} estimate for Sr dopants in La$_{2-x}$Sr$_x$CuO$_4$ (LSCO), is consistent with this picture [\onlinecite{Lee-Hone2020}].

	\section{ $\mathbf{T}$-matrix approach: intermediate scattering and unitary limit}
	
	After introducing the continuum Green's function approach, we continue to explore intermediate scattering and the strong scattering limit. 

\begin{figure}[t]
		\centering
		\includegraphics[width=1\linewidth]{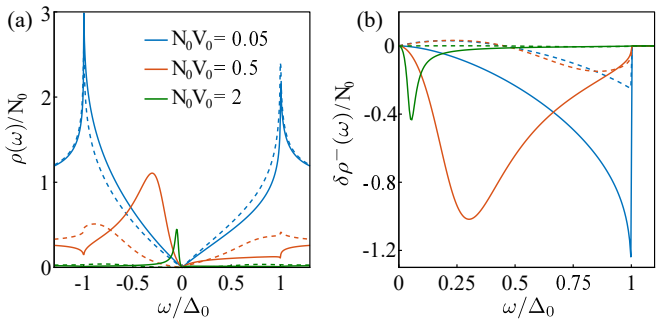}		
		\caption{Calculated (a) LDOS and (b) $\delta\rho^-_{d,s}(\omega)$ at the impurity site for different impurity potential strengths within the $T$-matrix approximation as a function of frequency, assuming a parabolic band dispersion and a $d$-wave $\Delta_0\cos(2\phi)$ gap (solid curves) or a sign-preserving $s$-wave $\Delta_0|\cos(2\phi)|$ gap (dashed curves).}	
		\label{Fig:BS_DOS}
	\end{figure}

\begin{figure}[t]
		\centering
		\includegraphics[width=0.5\textwidth]{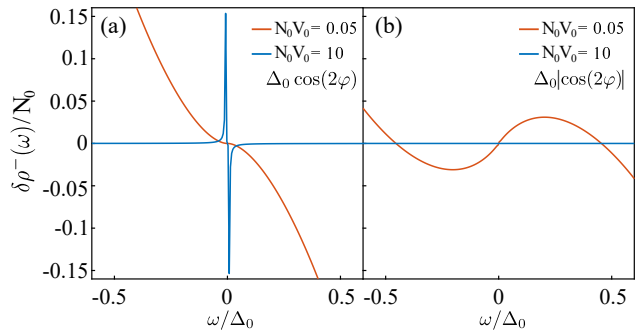}		
		\caption{Calculated $\delta\rho^-(\omega)$ for (a) $d$-wave [Eq.~(\ref{Eq:Limits1})] and (b) sign-preserving $s$-wave [Eq.~(\ref{Eq:Limits2})] gaps in the weak- (orange) and unitary- (blue) scattering  limits.} 	
		\label{Fig:Limits}
	\end{figure}

	In this case, the LDOS and its correction due to impurity scattering can be obtained within the $T$-matrix approach using 
	\begin{align}
	\delta \rho(\qv,\omega)=-\frac{1}{\pi}\sum_{\kk}\text{Im}\text{Tr}\frac{(\tau_0+\tau_3)}{2}\hat{G}_\kk(\omega)\hat{T}(\omega)\hat{G}_{\kk+\qv}(\omega),
	\label{eq:tmatrix}
	\end{align}
	with $\hat{T}(\omega)=(\hat{V}^{-1}-\sum_\textbf{k}\hat{G}_\textbf{k}(\omega))^{-1}$. Similar to the previous sections, we compare the results for the simplified parabolic dispersion and the lattice-based tight-binding Hamiltonian with momentum-dependent gaps. In Fig.~\ref{Fig:BS_DOS}, we present the LDOS and $\delta\rho^-_{d,s}(\omega)$ as measured at the impurity site using a parabolic band dispersion. For this case we used Green's functions which are given by Eqs.(\ref{Eq:SimpleGwd}) and (\ref{Eq_SimpleGws}) and different impurity potential strengths $V_0$, and used these in Eq.~(\ref{eq:tmatrix}).  The solid and dashed curves correspond to $\Delta_0\cos(2\phi)$ ($d$-wave) and $|\Delta_0\cos(2\phi)|$ (sign-preserving $s$-wave) gaps, respectively.
	Observe that with increasing potential an impurity bound state in the LDOS develops for the $d$-wave gap, whereas for the $s$-wave case the signal intensity gets simply suppressed. The shape of the bound state is then reflected in $\delta\rho^-_{d}(\omega)$ shown in Fig.~\ref{Fig:BS_DOS}(b). Consistent with the HAEM proposal in the Born limit, $\delta\rho^-_{d}(\omega)$ displays \emph{even} behavior in the frequency interval $0<\omega\leq \Delta_0$; however, the intensity maximum, which was located at coherence peak energy in the Born limit, now follows the bound state energy. The $\delta\rho^-_{s}(\omega)$ signal is \emph{odd} in frequency, and its intensity decreases for larger impurity potentials and converges to zero in the unitary limit. Analytically, one finds for the $d$-wave case at $\omega\ll \Delta_0$
	\begin{align}
	\delta\rho^-_d(\omega)&\approx
	\begin{cases}
	-\frac{8V_0N_0^2}{\pi}\text{sgn}(\omega)\frac{\omega^2}{\Delta_0^2}\ln\left(\frac{4\Delta_0}{|\omega|}\right)\ & |V_0| \rightarrow 0, \\
	-\frac{2\text{sgn}(\omega)}{N_0^2}\frac{ \frac{1}{V_0^3} }{ \frac{\omega^2}{\Delta_0^2} \ln\left(\frac{\Delta_0}{|w| }\right)^3} &  |V_0| \rightarrow \infty.\label{Eq:Limits1}
	\end{cases}.
	\end{align}
	The limit $V_0\rightarrow 0$ yields Eq.~(\ref{drho(i)Approx}), whereas  as $V_0\rightarrow \infty$ the signal is suppressed by $1/V_0^3$ everywhere except for $|\omega|/\Delta_0\rightarrow 0$, which leads a finite negative (positive) peak at $\omega\searrow 0$ ($\omega\nearrow 0$). 
	In the sign-preserving $s$-wave case for $\omega\ll \Delta_0$,
	\begin{align}
	\delta\rho^-_s(\omega)&\approx
	\begin{cases}
	\frac{8V_0N_0^2}{\pi}\left[\frac{\omega}{\Delta_0}-\text{sgn}(\omega)\frac{\omega^2}{\Delta_0^2}\ln\left(\frac{4\Delta_0}{|\omega|}\right)\right] & |V_0| \rightarrow 0, \\
	0 &  |V_0| \rightarrow \infty.\label{Eq:Limits2}
	\end{cases}
	\end{align}
	For $V_0\rightarrow 0$, one recovers Eq.~(\ref{drho(ii)Approx}) and for $V_0\rightarrow \infty$,  $\delta\rho^-_s(\omega)\rightarrow0$ as no bound state occurs at low energies. We plot the results of the analytical calculations in Fig.~\ref{Fig:Limits}.

	At this point one concludes that the main feature of the HAEM method---{\it i.e.}, \emph{even} or \emph{odd} behavior of the $\delta \rho^- (\omega)$ as a function of frequency in the interval $0<\omega\leq \Delta_0$ for sign-changing and sign-preserving gaps, respectively---also holds for nodal  superconductors from the Born limit to the unitary limit, with the only difference that the intensity maximum in $\delta \rho^-_d(\omega)$ of the latter is located at the bound state energy and not at $\omega=\Delta_0$. This is in contrast to the nodeless sign-changing $s^{\pm}$-wave gap in iron-based superconductors where the bound-state energy can be well separated from the structure in $\delta \rho^-(\omega)$ and the maximum of the latter still occurs at superconducting gap energy [\onlinecite{AltenfeldLiFeAs2018}]. There is also the difficulty in clarifying the precise origin of the bound state peak in unconventional nodal $d$-wave superconductors due to the possibility that a magnetic moment may be generated near a formally nonmagnetic impurity  [\onlinecite{Gabay2009}]. This represents a certain difficulty in using HAEM in the unitary limit.  
	
\begin{figure}[t]
		\centering
		\includegraphics[width=1\linewidth]{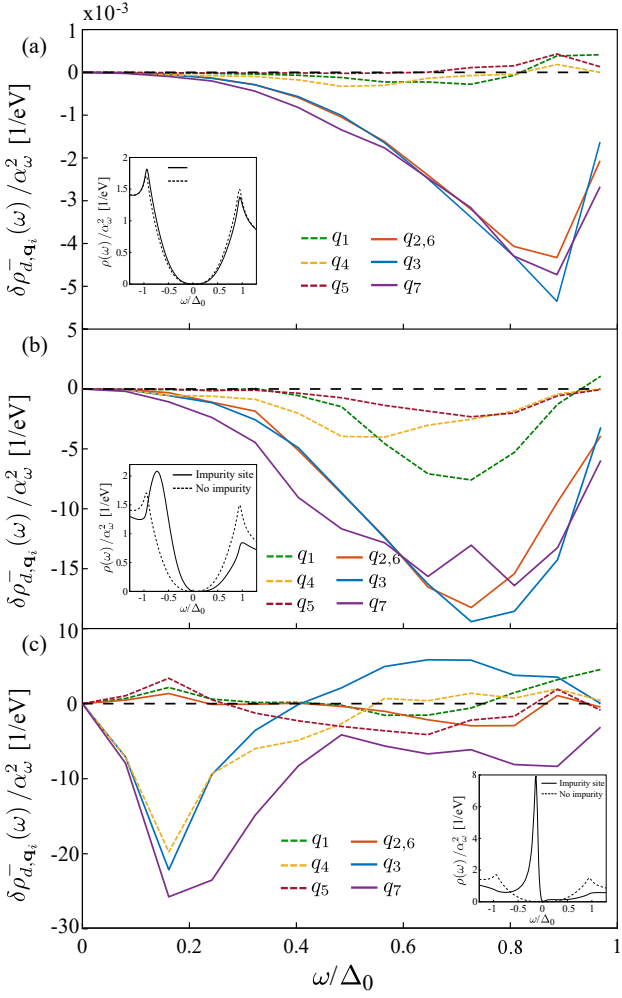}
		\caption{ $\delta\rho^-_d(\qv,\omega)/\alpha^2_\omega$ in units of [states/eV/spin] obtained using the full $T$-matrix approach and continuum Green's functions integrated in vicinity of each octet vectors for (a) weak ($V_0=-20$ meV), (b) intermediate ($V_0=-150$ meV), and (c) strong ($V_0=-1000$ meV) impurity potentials, the last of which causes a bound state at $\omega/\Delta_0\approx 0.16$. Insets show the LDOS as measured on the impurity site (solid curves) and far from impurity site (dashed curves) for the impurity potentials in (a), (b), and (c), respectively.}	
		\label{Fig:Strong_vs_Weak_Comparison}
	\end{figure}

\begin{figure}[t]
		\centering
		\includegraphics[width=1\linewidth]{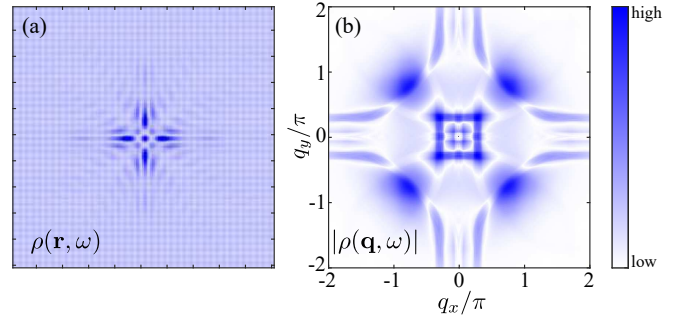}
		\caption{A real-space conductance map at negative bias with a $V_0=-1000$ meV impurity located at the center of the image as obtained from Eqs.~(\ref{Eq:BandDispersion})-(\ref{Eq:ContinuumLDOS}). (b) shows the Fourier transform of (a). }	
		\label{Fig:RealSpace_FourierSpace}
	\end{figure}

	In Fig.~\ref{Fig:Strong_vs_Weak_Comparison}(a)-\ref{Fig:Strong_vs_Weak_Comparison}(c), using the continuum description of Eqs.~(\ref{Eq:BandDispersion})-(\ref{Eq:ContinuumLDOS}), in which we replaced $\hat{V}\rightarrow \hat{T}(\omega)$, we compare the Fourier-transformed and $\qv_i$-integrated LDOS $\delta\rho^-_{d}(\qv,\omega)$ from the weak to intermediate and strong (unitary) impurity scattering regimes, respectively. The corresponding real-space conductance map for the unitary scattering limit is also shown in Fig.~\ref{Fig:RealSpace_FourierSpace}.
	Observe that the weak-scattering case (a) is consistent with the results of the Born limit and confirms the expectations from the HAEM approach. With increasing strength of the scattering potential the overall intensity of the $\delta\rho^-_{d,\qv_i}(\omega)$ for the sign-changing and sign-preserving octet wave vectors become similar in magnitude, yet also for intermediate scattering the corresponding even and odd frequency dependencies can be clearly identified [see Fig.~\ref{Fig:Strong_vs_Weak_Comparison}(b)]. This is also the case for the nodeless sign-changing $s$-wave gap [\onlinecite{AltenfeldLiFeAs2018}]. The situation changes dramatically for the unitary scattering limit shown in Fig.~\ref{Fig:Strong_vs_Weak_Comparison}(c). Here we also find $\delta\rho^-_{d,\qv_{3,7}}(\omega)$ has a maximum intensity near the bound state energy. Within the same energy region $\delta\rho^-_{d,\qv_{1,5}}(\omega)$ is less intense and and changes sign before reaching bound state energy. Both observations are qualitatively consistent with HAEM's theory with exception of $\delta\rho^-_{d,\qv_{3}}(\omega)$ changing sign at higher energies. Moreover,  $\delta\rho^-_{d,\qv_{4}}(\omega)$ follows the $\qv_3$-integrated signal, which is at odds with  the fact that $\qv_{4}$ is a sign-preserving vector. In addition, at low energies, the  $\qv_{2,6}$-integrated signal follows the $\qv_1$-integrated one and is much less intense than the $\qv_3$ and $\qv_7$ signal. We conclude that the existence of the bound state at low energies blurs the boundaries between sign-changing and sign-preserving scattering vectors within HAEM approach and redistributes the spectral weight of each by introducing new relevant scattering events. Consequently, it affects the momentum-resolved information from the octet vectors, which could be extracted from $\delta\rho^-_{d,\qv_i}(\omega)$. This strongly suggests that strong impurities in the unitary scattering limit have to be avoided within the HAEM approach.
	
\begin{figure}[t]
		\centering
		\includegraphics[width=1\linewidth]{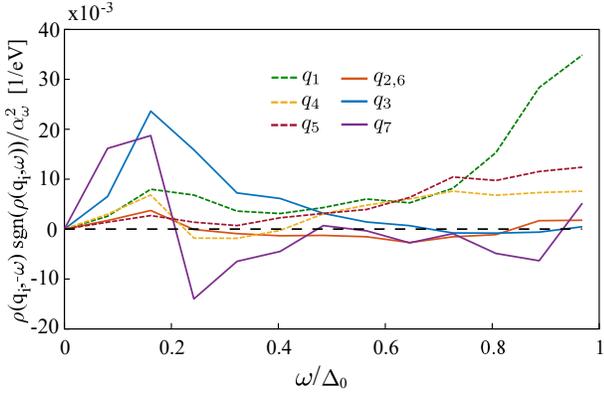}	
		\caption{The phase-sensitive quantity $\rho(\qv,-\omega)\times\text{sgn}(\rho(\qv,\omega))/\alpha^2_\omega$, introduced in Ref.~[\onlinecite{Chi2017a}], in units of [states/eV/spin] integrated in the vicinity of each octet vector for the strong impurity potential corresponding to Fig.~\ref{Fig:Strong_vs_Weak_Comparison}(c) ($V_0 = -1000$ meV). In theory all solid (dashed) curves are supposed to show negative (positive) signal. This is not the case everywhere and especially near the bound state energy.}
		\label{Fig:IBS_Quantity}
	\end{figure}

	Similar to  previous results [\onlinecite{Hirschfeld2015}], we conclude here that a clear distinction between sign-changing and sign-preserving gap functions is significantly suppressed or lost in the presence of strong scatterers, since most of the spectral weight is pulled into the bound state in the sign-changing case.  Furthermore, due to the nodal character of the gap, the formation of the bound state prevents {\bf q}-selective analysis of the sign-changing and sign-preserving superconducting gap regions. Thus the method is simply not reliable in the strong impurity limit. As mentioned in the Introduction an alternative phase-sensitive QPI analysis designed to work in the case of a strong impurity bound state (IBS) has been recently proposed by Chi \textit{et al}. [\onlinecite{Chi2017a}], and was apparently used successfully to analyze $d$-wave behavior in the presence of a weak scatterer in BSCCO [\onlinecite{Gu2019}].  Our calculations, however, also show that this method does not produce unambiguous results if the realistic continuum approach is followed (see Fig.~\ref{Fig:IBS_Quantity}), and is subject to the same caveats in the strong-impurity limit as the HAEM approach is.
	
	\section{HAEM for multiband $\mathbf{d}$-wave gap and three-dimensional electronic structure }
	
	The QPI technique as a probe to determine the momentum dependence of the superconducting gap has been also shown to work for the multiband  unconventional $d$-wave superconductors with a significant $k_z$ dispersion of the electronic structure such as CeCoIn$_5$ [\onlinecite{Akbari2011},\onlinecite{Allan2013}]. In this regard, it is interesting to check whether the HAEM approach would also work in this case. Instead of looking at CeCoIn$_5$, we consider the recently discovered infinite-layer Sr-doped $\text{NdNiO}_2$ thin film superconductors grown on SrTiO$_3$ [\onlinecite{Li2019_Ni_Discovery},\onlinecite{Li2020_ScNiDome}]. According to recent theoretical analyses [\onlinecite{NickelRhonny2020,Werner2020SpinFreezing,Sakakibara2019NiModel_dWaveFLEX}], infinite layer nickelates possess a strong tendency towards a robust $d_{x^2-y^2}$ symmetry of the superconducting gap as a dominant superconducting instability despite the sizable three-dimensionality of the electronic structure due to admixing of the Nd 5$d_{z^2}$ states at the Fermi level in addition to the mostly two-dimensional $d_{x^2-y^2}-$ character of the Ni 3$d$ states. Therefore the HAEM approach would be useful to apply here to verify the symmetry of the superconducting gap. In particular, we employ the electronic structure of Sr-doped $\text{NdNiO}_2$ using the three-dimensional three-orbital tight-binding Hamiltonian from Ref.~[\onlinecite{NickelRhonny2020}],
	\begin{align}
	H=\sum_{\kk,\sigma}\Psi^\dagger_{\kk,\sigma}h(\kk)\Psi_{\kk,\sigma}\label{Eq:NickelHamiltonian},
	\end{align} 
	where $\Psi^\dagger_{\kk,\sigma}=[c^\dagger_{z^2,\sigma}(\kk),c^\dagger_{xy,\sigma}(\kk),c^\dagger_{x^2-y^2,\sigma}(\kk)]$ contains fermionic operators with spin $\sigma$ which create particles in Nd 5$d_{z^2}$, Nd 5$d_{xy}$, and Ni 3$d_{x^2-y^2}$ orbitals, respectively. The matrix elements and hopping parameters of $h(\kk)$ are given in Appendix C. We parametrize the superconducting pairing matrix as $\hat{\Delta}_\kk^{d(s)}=\text{diag}(\Delta^{d(s)}_{z^2}(\kk),\Delta^{d(s)}_{xy}(\kk),\Delta^{d(s)}_{x^2-y^2}(\kk))$, where $\Delta^d_\alpha(\kk)=\frac{\Delta_{\alpha 0}}{2}(\cos(k_x)-\cos(k_y))$, and for simplicity we assume the same size of the superconducting gap at each orbital $\Delta_{\alpha 0}/|t^x_{33}| = 0.1329$. 
	\begin{figure}[t]
		\centering
		\includegraphics[width=1\linewidth]{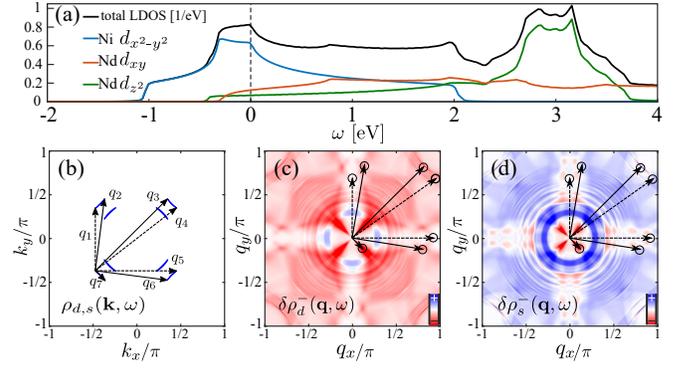}		
		\caption{(a) Normal-state DOS and its contributions from each of the orbitals. (b) Contours of constant quasiparticle energy in the superconducting state at $k_z=0$. Scattering vectors $q_1$-$q_7$ connect tips of the banana-shaped pockets resulting from Ni $d_{x^2-y^2}$ orbitals. (c) $\delta\rho^-_d(\qv,\omega)$ and (d) $\delta\rho^-_s(\qv,\omega)$ QPI pattern at $q_z=0$ and $\omega/\Delta_{x^2-y^2}=0.16$ with the impurity in the center of the image.}	
		\label{Fig:NickelMaps}
	\end{figure}

\begin{figure}[t]
		\centering
		\includegraphics[width=1\linewidth]{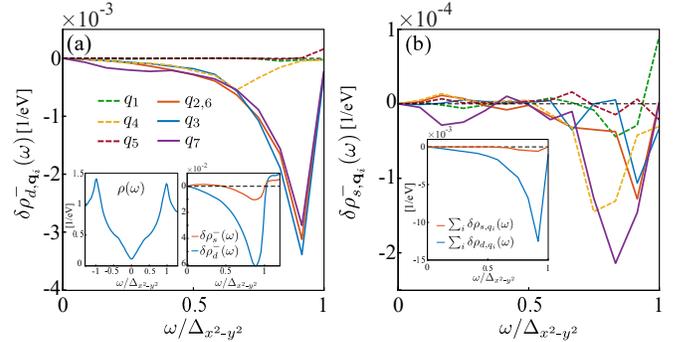}		
		\caption{Antisymmetrized correction to LDOS in units of [states/eV/spin] integrated in the vicinity of each octet vector at $k_z=0$ [see Fig.~\ref{Fig:NickelMaps}(c) and \ref{Fig:NickelMaps}(d)] for (a) $d$-wave and (b) sign-preserving $s$-wave pairing symmetry gaps. Solid (dashed) curves correspond to signals that stem from sign-changing (sign-preserving) octet scattering vectors. Inset (a) left:  momentum-integrated LDOS $\rho(\omega)$. Inset (a) right: antisymmetrized correction $\delta\rho^-_{d(s)}(\omega)$ integrated over all momenta, including $q_z$. Inset (b) shows summed results from (a) and (b) plotted within the same energy scale.}	
		\label{Fig:Nickel_DOS}
	\end{figure}

	In Fig.~\ref{Fig:NickelMaps}(a), we show the normal-state DOS and its orbital contributions as obtained from Eq.~(\ref{Eq:NickelHamiltonian}). Near the Fermi energy, the DOS consists mostly of contributions from the almost two-dimensional Ni 3$d_{x^2-y^2}$-states as well as Nd 5$d_{z^2}$ and 5$d_{xy}$ orbitals, in agreement with previous results [\onlinecite{Nomura2019_Ni_3_OrbFS}]. We find that the low-energy features of $\rho(\qv,\omega)$ in the superconducting state are almost entirely dominated by the gap on the $d_{x^2-y^2}$ orbital which sets the gap scale that determines the position of the coherence peaks [see inset of Fig.~\ref{Fig:Nickel_DOS}(a)]. We observe that a variation in the $\Delta_{z^20}/\Delta_{xy0}$ parameter size and/or their relative sign hardly effects the LDOS; we thus expect similar HAEM results for the nickelates as for the cuprates. The momentum-integrated $\delta\rho^-_{d,s}(\omega)$ computed for all momenta (also including $q_z$) within the Born limit are shown in the inset of Fig.~\ref{Fig:Nickel_DOS}(a) (right inset) and obey the expected \emph{even} and \emph{odd} frequency-dependent behavior for sign-changing and sign-preserving gaps, respectively. This shows that HAEM can be also used in the three-dimensional case as a probe of gap sign. 
	%  which indicates the sign-changing gap in general. 
	%

	In contrast to the single-band model typical for cuprates, in the infinite layer nickelate, the Nd 5$d_z^2$ orbital also contributes to the states at the Fermi level and forms a three-dimensional pocket centered near the $\Gamma$ point of the Brillouin zone, which enlarges the space of potential wave vectors involved in the scattering processes. However, similar to CeCoIn$_5$ [\onlinecite{Allan2013}], we find that only those scattering vectors that connect tips of the banana-shaped pockets resulting from the Ni $d_{x^2-y^2}$ orbitals dominate QPI spectra. We label them $q_1$ to $q_7$ as for the cuprates [see Figs.~\ref{Fig:NickelMaps}(b)-\ref{Fig:NickelMaps}(d)]. As in the single band case, the $\qv_i$-integrated $\delta\rho^-_d(\qv,\omega)$ for the $k_z=0$ cut shows an \emph{even} (\emph{odd}) behavior for the sign-changing $q_{2,6}$, $q_3$ and $q_7$ (sign-preserving $q_{1}$, $q_4$ and $q_5$) vectors, as depicted in Fig.~\ref{Fig:Nickel_DOS} (a). Furthermore, the $\qv_i$-integrated $\delta\rho^-_s(\qv,\omega)$ for the sign-preserving gap behaves \emph{odd} in frequency for all seven vectors. Both results are in agreement with the HAEM proposal. The inset in Fig.~\ref{Fig:Nickel_DOS}(b) shows the {\bf q$_i$}-summed results of (a) and (b) on the same energy scale done in a similar fashion as in previous sections, which is in good qualitative agreement with the fully momentum-integrated results.  This indicates that HAEM with weak and intermediate impurity scattering could be also applicable to quasi-two-dimensional unconventional nodal superconductors with sizable electronic dispersion along the $k_z$ direction. Our results can be tested experimentally in the doped NdNiO$_2$ system.
	
	\section{Conclusions}
	An important problem in the field of unconventional superconductivity is to determine the phase or sign structure of the order parameter.  In the current paper, we have extended the phase-sensitive QPI method of Ref.~[\onlinecite{Hirschfeld2015}] (HAEM) to systems where the gap is strongly momentum-dependent with nodal structure,  and we have treated both single-band and multiband systems using local and continuum Green's function approaches. In particular, we showed that the existence of a sign change of the gap is reflected qualitatively in the $\bf q$-integrated antisymmetrized conductance from Born to the unitary limits of the impurity scattering and is therefore a universal hallmark of the sign-changing gap. In addition,  in the Born limit and for intermediate scattering strength, the selective $\bf q$-integration around corresponding scattering peak positions in $\bf q$ space (the so-called ``octet QPI peaks'' in the case of $d$-wave superconductors) exhibits the HAEM separation between sign-preserving and sign-changing scattering processes. At the same time, neither the HAEM nor the  IBS methods may be used q selectively in the unitary scattering limit due to the destructive interference of the bound state peak. This is in contrast to the nodeless sign-changing gap superconductors where the HAEM and IBS could be still used in the unitary scattering limit [\onlinecite{Hirschfeld2015}]. We have discussed the extension of the HAEM approach to quasi-two-dimensional superconductors with sizable $k_z$ dispersion and proposed the application of HAEM to the recently discovered infinite-layer nickelate superconductor NdNiO$_2$. We believe that phase-sensitive momentum-resolved quasiparticle interference analyses of this type will be a powerful tool to identify unconventional superconductors and to map their gap phase structure in the future.
	
	\section{Acknowledgements}
	The authors are grateful for valuable discussions with P. Choubey and  Z. Du and to D. Altenfeld for his work in the early stages of this project. M.A.S. was supported by the IARPA SuperTools program. P.J.H. received partial support from NSF-DMR-1849751.

	\bibliography{QPIBib}
	
	\appendix
	
	\begin{widetext}
		
		\section{Momentum-integrated Green's function\label{Sec:MomentumIntegratedGreensfunction}}
		
		Assuming a parabolic dispersion, the momentum-integrated Green's function is given by
		\begin{align}
		\hat{G}^{s(d)}(\omega)=N_0\int_{0}^{2\pi}\frac{d\varphi}{2\pi}\frac{-i\pi \text{sgn}(\omega)\left(\begin{array}{cc}
			\omega & \Delta^{s(d)}_\kk \\ 
			\Delta^{s(d)}_\kk & \omega
			\end{array} \right)}{\sqrt{\omega^2-\Delta_0^2\cos^2(2\varphi)+i\delta\text{sgn}(\omega)}}\label{Eq:EnergyIntegrated}.
		\end{align}	
		When integrating over the polar angle, the off-diagonal elements in Eq.~(\ref{Eq:EnergyIntegrated}) vanish for a $d$-wave gap, but remain finite for the $s$-wave gap, and we find
		\begin{align}
		\hat{G}^d(\omega)&=-2iN_0K\left[\frac{\Delta_0}{\omega}\right]\tau_0,\\
		\hat{G}^s(\omega)&=-2iN_0\Bigg[ K\left[\frac{\Delta_0}{\omega}\right]\tau_0-\left(\ln\left(\frac{\sqrt{|\omega^2-\Delta_0^2|}}{\Delta_0+|\omega|}\right)\text{sgn}(\omega)+\frac{i\pi}{2}\theta\left[\Delta_0^2-\omega^2\right]\right)\tau_1\Bigg],
		\end{align}
		where $K\left[\frac{\Delta_0}{\omega}\right]$ is the elliptic function of the first kind:
		\begin{align}
		K\left[\frac{\Delta_0}{\omega}\right]=\int_{0}^{\pi/2}d\theta\frac{1}{\sqrt{1-\frac{\Delta_0^2}{\omega^2}\sin^2(\theta)+i\delta\text{sgn}(\omega)}}.
		\end{align}

		\section{Calculating $ \delta\rho(\kk_F,\qv_i,\omega) $\label{Suppl:AnalyticalPoints}}
		We calculate $\delta\rho^-(\kk,\qv,\omega)$ and evaluate it at $\kk=\kk_F$ and $\qv=\qv_i$, where $\qv_i$ is an octet vector. At these particular points we have $\epsilon_{\kk_F}=\epsilon_{\kk_F+\qv_i}=0$ and $\Delta_{\kk_F+\qv_i}=\pm\Delta_{\kk_F}$, with the ``$+$'' solution for sign-preserving $(\qv_1,\qv_4,\qv_5)$ and the ``$-$'' solution for sign-changing $(\qv_{2/6},\qv_3,\qv_7)$. For $\delta\rho(\kk_F,\qv_i,\omega)=-\frac{1}{\pi}\text{ImTr}\frac{\tau_0+\tau_3}{2}\hat{G}_{\kk_F}(\omega)V_0\tau_3\hat{G}_{\kk+\qv_i}(\omega)$, one finds
		
		\begin{align}
		\delta\rho(\kk_F,\qv_i,\omega)&=
		-\frac{V_0}{\pi}\text{Im}\Bigg(\frac{\omega+i\delta}{(\omega+i\delta)^2-|\Delta_{\kk_F}|^2}\frac{\omega+i\delta}{(\omega+i\delta)^2-|\Delta_{\kk_F+\qv_i}|^2}
		-\frac{\Delta_{\kk_F}}{(\omega+i\delta)^2-|\Delta_{\kk_F}|^2}\frac{\Delta_{\kk_F+\qv_i}}{(\omega+i\delta)^2-|\Delta_{\kk_F+\qv_i}|^2}\Bigg)\\
		&=
		-\frac{V_0}{\pi}\text{Im}\Bigg(\frac{(\omega+i\delta)^2-\Delta_{\kk_F}\Delta_{\kk_F+\qv_i}}{\left[(\omega+i\delta)^2-|\Delta_{\kk_F}|^2\right]^2}
		\Bigg)\label{Eq:StaringPoint1}\\
		&=
		-\frac{V_0}{\pi}\text{Im}\Bigg(\frac{(\omega+i\delta)^2-\Delta_{\kk_F}\Delta_{\kk_F+\qv_i}}{\left[(\omega+i\delta)^2-|\Delta_{\kk_F}|^2\right]^2}
		\times\frac{\left[(\omega-i\delta)^2-|\Delta_{\kk_F}|^2\right]^2}{\left[(\omega-i\delta)^2-|\Delta_{\kk_F}|^2\right]^2}\Bigg)\label{Eq:StaringPoint2}\\
		&=-\frac{V_0}{\pi}\text{Im}\Bigg(
		\frac{((\omega+i\delta)^2-\Delta_{\kk_F}\Delta_{\kk_F+\qv_i})\left[(\omega-i\delta)^2-|\Delta_{\kk_F}|^2\right]^2}
		{\Big|\left[(\omega+i\delta)^2-|\Delta_{\kk_F}|^2\right]^2\Big|^2},
		\Bigg),\label{Eq:StaringPoint3}
		\end{align}	
		where by expanding with the complex conjugate of the denominator in Eq.~(\ref{Eq:StaringPoint1}), the imaginary part is isolated in the numerator of Eq.~(\ref{Eq:StaringPoint3}), which contains the crucial phase-sensitive term $\Delta_{\kk_F}\Delta_{\kk_F+\qv_i}$.
		The denominator in Eq.~(\ref{Eq:StaringPoint3}) is even in $\omega$ and reads
		\begin{align}
		\Big|\left[(\omega+i\delta)^2-|\Delta_{\kk_F}|^2\right]^2\Big|^2=\left[\left(\Delta_{\kk_F}^2-\omega^2)^2+\delta^4+2\delta^2(\omega^2+\Delta_{\kk_F}^2\right)\right]^2.\label{Eq:Den}
		\end{align}
		For the numerator, we find
		\begin{align}
		&-\frac{V_0}{\pi}\text{Im}\left\{((\omega+i\delta)^2-\Delta_{\kk_F}\Delta_{\kk_F+\qv_i})\left[(\omega-i\delta)^2-|\Delta_{\kk_F}|^2\right]^2\right\}\\
		=&\begin{cases}
		\frac{2\omega\delta V_0}{\pi}\left[\left(\Delta_{\kk_F}^2-\omega^2\right)^2+\delta^4+2\delta^2\left(\omega^2+\Delta_{\kk_F}^2\right)\right]\quad &\text{for}\enspace \Delta_{\kk_F+\qv_i}=+\Delta_{\kk_F}\\\\
		\frac{2\omega\delta V_0}{\pi}\left[\left(\Delta_{\kk_F}^2+\omega^2\right)^2-4\Delta_{\kk_F}^4+\delta^4+2\delta^2\left(\omega^2-\Delta_{\kk_F}^2\right)\right]\quad &\text{for}\enspace\Delta_{\kk_F+\qv_i}=-\Delta_{\kk_F}
		\end{cases}\label{Eq:Cases}.
		\end{align}	
		Both cases in Eq.~(\ref{Eq:Cases}) are odd in frequency and thus contribute to $\delta\rho^-(\omega)$. Combining Eq.~(\ref{Eq:Den}) and  Eq.~(\ref{Eq:Cases}) yields
		\begin{align}
		\delta\rho^-_{++}(\omega)&=\frac{4V_0}{\pi}\frac{\omega\delta }{\left[\left(\Delta_{\kk_F}^2-\omega^2)^2+\delta^4+2\delta^2(\omega^2+\Delta_{\kk_F}^2\right)\right]}\label{Eq:RhoMinusSignPreserve},\\
		\delta\rho^-_{+-}(\omega)&=\frac{4V_0}{\pi}\omega\delta\frac{\left(\Delta_{\kk_F}^2+\omega^2\right)^2-4\Delta_{\kk_F}^4+\delta^4+2\delta^2\left(\omega^2-\Delta_{\kk_F}^2\right)}{\left[\left(\Delta_{\kk_F}^2-\omega^2)^2+\delta^4+2\delta^2(\omega^2+\Delta_{\kk_F}^2\right)\right]^2}\label{Eq:RhoMinusSignChange},
		\end{align}
		for sign-preserving ($++$) and sign-changing ($+-$) scattering, respectively.
		In order to extract information out of Eqs.~(\ref{Eq:RhoMinusSignPreserve}) and  (\ref{Eq:RhoMinusSignChange}) one needs to investigate their behavior for $\delta\rightarrow0$. To do so we remind the reader that the LDOS in the absence of scattering is proportional to the imaginary part of the Green`s function $\rho(\omega)\sim\pi\sum_\kk \delta(\omega^2-\epsilon_{\kk}^2-\Delta_\kk^2 )$ with $\delta(x)$ the Dirac delta function.	Since in our approximation $\epsilon_{\kk_F}=0$, this expression, and also Eqs.~\ref{Eq:RhoMinusSignPreserve} and \ref{Eq:RhoMinusSignChange}, will have a nonzero value only in the limit $\lim_{\omega\nearrow\Delta_{\kk_F}}$. Hence, we introduce the small quantity $\epsilon\ll \Delta_{\kk_F}$ and write
		\begin{align}
		\omega\equiv\Delta_{\kk_F}-\epsilon\label{Eq:Limit},
		\end{align}
		which gives $\Delta_{\kk_F}^2-\omega^2\approx 2\Delta_{\kk_F}\epsilon$ and $\Delta_{\kk_F}^2+\omega^2\approx 2\Delta_{\kk_F}+2\Delta_{\kk_F}\epsilon$. Inserting Eq.~\ref{Eq:Limit} into denominators of Eq.~(\ref{Eq:RhoMinusSignPreserve}) and (\ref{Eq:RhoMinusSignChange}) and expanding to linear order in $\delta$ gives
		\begin{align}
		\delta\rho^-_{++}(\omega)&=\frac{V_0}{\pi}\frac{\omega\delta}{\Delta_{\kk_F}^2(\epsilon^2+\delta^2)}\approx\frac{V_0}{\pi}\frac{\omega\delta}{\Delta_{\kk_F}^2\epsilon^2}\label{Eq:ResultSignPreserving}\\
		\delta\rho^-_{+-}(\omega)&=\frac{V_0}{\pi}\frac{-\omega\delta}{\Delta_{\kk_F}^2(\epsilon^2+\delta^2)}\frac{2\Delta_{\kk_F}\epsilon}{\epsilon^2+\delta^2}
		\approx-\frac{V_0}{\pi}\frac{\omega\delta}{\Delta_{\kk_F}^2\epsilon^2}\frac{2\Delta_{\kk_F}}{\epsilon}.\label{Eq:ResultSignChanging},
		\end{align} 
		which yields the ratio $\delta\rho^-_{+-}(\omega)/\delta\rho^-_{++}(\omega)\approx-2\Delta_{\kk_F}/\epsilon$.
		At this point one cannot make a statement regarding possible sign changes for sign-preserved scattering in Eq.~(\ref{Eq:ResultSignPreserving}) which requires contributions from $\kk\neq\kk_F$. However, while $ \delta\rho^-_{++}(\omega) $ is positive, $ \delta\rho^-_{+-}(\omega) $ is negative and larger in magnitude by a factor $2\Delta_{\kk_F}/\epsilon$, revealing a clear hierarchy in intensity distribution.

		\section{Tight-Binding Parameters for NdNiO$\mathbf{_2}$}
		Tight-binding fitting parameters in units of eV and matrix elements of Eq.~(\ref{Eq:NickelHamiltonian}) adopted from Ref.~[\onlinecite{NickelRhonny2020}]:
		\begin{align}
		\mu&=6.5814\nn\\	
		\epsilon_1&=8.9506,\enspace\epsilon_2=9.0277,\enspace\epsilon_3=6.8979,\enspace\nn\\	
		t^x_{11}&=-0.3870,\enspace t^{xy}_{11}=0,\enspace t^{xx}_{11}=0.034,\enspace t^z_{11}=-0.8591,\enspace t^{xz}_{11}=0.0107,\enspace t^{xyz}_{11}=0.025,\enspace t^{zz}_{11}=0.0904\nn\\		
		t^x_{22}&=0.3202,\enspace t^{xy}_{22}=-0.0467,\enspace t^{xx}_{22}=0.0367,\enspace t^z_{22}=0.3216,\enspace t^{xz}_{22}=-0.1438,\enspace t^{xyz}_{22}=0.0496,\enspace\nn\\ 
		t^{zz}_{22}&=-0.0327,\enspace\nn\\
		t^{xxz}_{22}&=-0.0209,\enspace t^{xxy}_{22}=-0.0198,\enspace t^{xxyz}_{22}=0.0164,\enspace t^{xxx}_{22}=0.012,\enspace\nn\\		
		t^{xy}_{12}&=0.0798,\enspace t^{xyz}_{12} =-0.0669,\enspace t^{xyzz}_{12}=0.0094,\enspace\nn\\		
		t^x_{33}&=-0.3761,\enspace t^{xy}_{33}=0.0844,\enspace t^{xx}_{33}=-0.0414,\enspace t^{xxy}_{33}=-0.0043,\enspace t^{xxyy}_{33}=0.003,\enspace t^z_{33}=-0.0368,\enspace\nn\\
		t^{xz}_{33}&=-0.0019,\enspace\nn\\	t^{xyz}_{33}&=0.0117,\enspace t^{zz}_{33}=0.008,\enspace\nn\\		
		t^{xxy}_{13}&=0.0219,\enspace\nn\\		
		t^{xxy}_{23}&=-0.0139,\enspace\nn\\		
		t^{xxy}_{11}&=0,\enspace\nn\\		
		\end{align}
		\begin{align}		
		h_{11}&=\epsilon_1-\mu+2t^x_{11}[\cos(k_x)+\cos(k_y)]+4t^{xy}_{11}\cos(k_x)\cos(k_y)+2t^{xx}_{11}[\cos(2k_x)+\cos(2k_y)]\nn\\
		&+4t^{xxy}_{11}[\cos(2k_x)\cos(k_y)+\cos(k_x)\cos(2k_y)]+2t^z_{11}\cos(k_z)+2t^{zz}_{11}\cos(2kz)\nn\\
		&+4t^{xz}_{11}\cos(k_z)[\cos(k_x)+\cos(k_y)]+8t^{xyz}_{11}\cos(k_x)\cos(k_y)\cos(k_z)\\\nn\\				
		h_{22}&=\epsilon_2-\mu+2t^x_{22}[\cos(k_x)+\cos(k_y)]+4t^{xy}22\cos(k_x)\cos(k_y)+2t^{xx}_{22}[\cos(2k_x)+\cos(2k_y)]\nn\\
		&+4t^{xxy}_{22}[\cos(2k_x)\cos(k_y)+\cos(k_x)\cos(2k_y)]+2t^z_{22}\cos(k_z)+2t^{zz}_{22}\cos(2k_z)\nn\\
		&+4t^{xz}_{22}\cos(k_z)[\cos(k_x)+\cos(k_y)]+8t^{xyz}_{22}\cos(k_x)\cos(k_y)\cos(k_z)+8t^{xxz}_{22}\cos(k_z)[\cos(2k_x)+\cos(2k_y)]\nn\\
		&+8t^{xxyz}_{22}\cos(k_z)[\cos(2k_x)\cos(k_y)+\cos(k_x)\cos(2k_y)]+2t^{xxx}_{22}[\cos(3k_x)+\cos(3k_y)]\\\nn\\					
		h_{33}&=\epsilon_3-\mu+2t^x_{33}[\cos(k_x)+\cos(k_y)]+4t^{xy}_{33}\cos(k_x)\cos(k_y)+2t^{xx}_{33}[\cos(2k_x)+\cos(2k_y)]\nn\\
		&+4t^{xxy}_{33}[\cos(2k_x)\cos(k_y)+\cos(k_x)\cos(2k_y)]+4t^{xxyy}_{33}\cos(2k_x)\cos(2k_y)+2t^z_{33}\cos(k_z)+2t^{zz}_{33}\cos(2k_z)\nn\\
		&+4t^{xz}_{33}\cos(k_z)[\cos(k_x)+\cos(k_y)]+8t^{xyz}_{33}\cos(k_x)\cos(k_y)\cos(k_z)\\\nn\\				
		h_{12}&=-4t^{xy}_{12}\sin(k_x)\sin(k_y)-8t^{xyz}_{12}\sin(k_x)\sin(k_y)\cos(k_z)-8t^{xyzz}_{12}\sin(k_x)\sin(k_y)\cos(2k_z)\\		
		h_{13}&=-8t^{xxy}_{13}\cos\left(\frac{k_z}{2}\right)\left[\sin\left(\frac{3kx}{2}\right)\sin\left(\frac{ky}{2}\right)-\sin\left(\frac{kx}{2}\right)\sin\left(\frac{3ky}{2}\right)\right]\\		
		h_{23}&=8t^{xxy}_{23}\cos\left(\frac{k_z}{2}\right)\left[ \cos\left(\frac{3k_x}{2}\right)\cos\left(\frac{k_y}{2}\right)-\cos\left(\frac{k_x}{2}\right)\cos\left(\frac{3k_y}{2}\right)\right]
		\end{align}
		
	\end{widetext}

\end{document}